\documentstyle[12pt]{article}
\def\PRL #1 #2 #3{{\sl Phys. Rev. Lett.} {\bf#1} (#2) #3}
\def\NPB #1 #2 #3{{\sl Nucl. Phys.} {\bf B#1} (#2) #3}
\def\NPBFS #1 #2 #3 #4{{\sl Nucl. Phys.} {\bf B#2} [FS#1] (#3) #4}
\def\CMP #1 #2 #3{{\sl Commun. Math. Phys.} {\bf #1} (#2) #3}
\def\PRD #1 #2 #3{{\sl Phys. Rev.} {\bf D#1} (#2) #3}
\def\PLA #1 #2 #3{{\sl Phys. Lett.} {\bf #1A} (#2) #3}
\def\PLB #1 #2 #3{{\sl Phys. Lett.} {\bf #1B} (#2) #3}
\def\JMP #1 #2 #3{{\sl J. Math. Phys.} {\bf #1} (#2) #3}
\def\PTP #1 #2 #3{{\sl Prog. Theor. Phys.} {\bf #1} (#2) #3}
\def\SPTP #1 #2 #3{{\sl Suppl. Prog. Theor. Phys.} {\bf #1} (#2) #3}
\def\AoP #1 #2 #3{{\sl Ann. of Phys.} {\bf #1} (#2) #3}
\def\PNAS #1 #2 #3{{\sl Proc. Natl. Acad. Sci. USA} {\bf #1} (#2) #3}
\def\RMP #1 #2 #3{{\sl Rev. Mod. Phys.} {\bf #1} (#2) #3}
\def\PR #1 #2 #3{{\sl Phys. Reports} {\bf #1} (#2) #3}
\def\AoM #1 #2 #3{{\sl Ann. of Math.} {\bf #1} (#2) #3}
\def\UMN #1 #2 #3{{\sl Usp. Mat. Nauk} {\bf #1} (#2) #3}
\def\FAP #1 #2 #3{{\sl Funkt. Anal. Prilozheniya} {\bf #1} (#2) #3}
\def\FAaIA #1 #2 #3{{\sl Functional Analysis and Its Application} {\bf
#1} (#2) #3}
\def\BAMS #1 #2 #3{{\sl Bull. Am. Math. Soc.} {\bf #1} (#2)
#3} \def\TAMS #1 #2 #3{{\sl Trans. Am. Math. Soc.} {\bf #1} (#2) #3}
\def\InvM #1 #2 #3{{\sl Invent. Math.} {\bf #1} (#2) #3}
\def\LMP #1 #2 #3{{\sl Letters in Math. Phys.} {\bf #1} (#2) #3}
\def\IJMPA #1 #2 #3{{\sl Int. J. Mod. Phys.} {\bf A#1} (#2) #3}
\def\AdM #1 #2 #3{{\sl Advances in Math.} {\bf #1} (#2) #3}
\def\RMaP #1 #2 #3{{\sl Reports on Math. Phys.} {\bf #1} (#2) #3}
\def\IJM #1 #2 #3{{\sl Ill. J. Math.} {\bf #1} (#2) #3}
\def\APP #1 #2 #3{{\sl Acta Phys. Polon.} {\bf #1} (#2) #3}
\def\TMP #1 #2 #3{{\sl Theor. Mat. Phys.} {\bf #1} (#2) #3}
\def\JPA #1 #2 #3{{\sl J. Physics} {\bf A#1} (#2) #3}
\def\JSM #1 #2 #3{{\sl J. Soviet Math.} {\bf #1} (#2) #3}
\def\MPLA #1 #2 #3{{\sl Mod. Phys. Lett.} {\bf A#1} (#2) #3}
\def\JETP #1 #2 #3{{\sl Sov. Phys. JETP} {\bf #1} (#2) #3}
\def\JETPL #1 #2 #3{{\sl  Sov. Phys. JETP Lett.} {\bf #1} (#2) #3}
\def\PHSA #1 #2 #3{{\sl Physica} {\bf A#1} (#2) #3}
\def\CQG #1 #2 #3{{\sl Class. Quantum Grav.} {\bf #1} (#2) #3}
\def\SJNP #1 #2 #3{{\sl Sov. J. Nucl. Phys. (Yadern.Fiz.)} {\bf #1} (#2) #3}
\def\a{\alpha}\def\b{\beta}\def\g{\gamma}\def\d{\delta}

\def\k{\kappa}
\def\Th{\Theta}\def\G{\Gamma}

\newcommand{\p}[1]{(\ref{#1})}
\begin{document}
\thispagestyle{empty}
\renewcommand{\thefootnote}{\fnsymbol{footnote}}
\begin{flushright}
Preprint DFPD 97/TH/19\\
ICTP IC/97/44\\
hep-th/9705064\\
May 1997
\end{flushright}

\bigskip
\begin{center}
{\large\bf Superbrane Actions and Geometrical Approach}
\footnote{Based on talks given by the authors at the Volkov Memorial
Seminar ``Supersymmetry and Quantum field Theory" (Kharkov, January 5--7,
1997)}

\vspace{1cm}
Igor Bandos$^1$\footnote{
On leave from NSC, Kharkov Institute of Physics and Technology,
Kharkov, Ukraine.\\
\hspace*{20pt}ibandos@ictp.trieste.it},
Paolo Pasti$^2$\footnote{paolo.pasti@pd.infn.it},
Dmitri Sorokin$^3$\footnote{dsorokin@kipt.kharkov.ua} and
Mario Tonin$^2$\footnote{mario.tonin@pd.infn.it}

\vspace{0.5cm}
$^1${\it International Centre for Theoretical Physics\\
34100. Trieste, Italy}

\bigskip
$^2${\it Universit\`a Degli Studi Di Padova,
Dipartimento Di Fisica ``Galileo Galilei''\\
ed INFN, Sezione Di Padova,
Via F. Marzolo, 8, 35131 Padova, Italia}

\bigskip
$^3${\it National Science Center,
Kharkov Institute of Physics and Technology,\\
Kharkov, 310108, Ukraine}

\vspace{1.cm}
{\bf Abstract}
\end{center}
We review a generic structure of
conventional (Nambu--Goto and Dirac--Born--Infeld--like) worldvolume
actions for the superbranes and show how it is connected through a
generalized action construction with a doubly supersymmetric geometrical
approach to the description of super--p--brane dynamics as embedding world
supersurfaces into target superspaces.

\renewcommand{\thefootnote}{\arabic{footnote}}
\setcounter{footnote}0
\bigskip
\bigskip
During last years Dmitrij Vasilievich Volkov actively studied geometrical
and symmetry grounds underlying the theory of supersymmetric extended
objects and we are happy to have been his collaborators in this work. One
of the incentives for this research was to understand the nature of an
important fermionic $\k$--symmetry of the target--superspace (or
Green--Schwarz) formulation of the superparticles and superstrings with
the aim to resolve the problem of its infinite reducibility, to relate the
Green--Schwarz and Ramond--Neveu--Schwarz formulation of superstrings
already at the classical level and to attack the problem of covariant
quantization of superstrings.  The $\k$--symmetry was conjectured to be a
manifestation of local extended supersymmetry (irreducible by definition)
on the world supersurface swept by a super--p--brane in a target
superspace. This was firstly proved for $N=1$ superparticles in three and
four dimensions \cite{stv} and then for $N=1$, $D=6,10$ superparticles
\cite{gs92}, $N=1$ \cite{hsstr}, $N=2$ \cite{gs2} superstrings, $N=1$
supermembranes \cite{tp93} and finally for all presently known
super--p--branes \cite{bpstv,bsv,hs1} in all space--time dimensions where
they exist.  In \cite{gs2rns} a twistor transform was applied to relate
the Green--Schwarz and the Ramond--Neveu--Schwarz formulation.

The approach to describing the
super--p--branes in this way is called the doubly supersymmetric
geometrical approach, since it essentially exploits the theory of
embedding world {\it super}surfaces into target {\it super}spaces. Apart
from having clarified the geometrical nature of $\k$--symmetry and having
made a substantial impact on the development of new methods of superstring
covariant quantization (see \cite{bzst,berko} and references therein),
the doubly supersymmetric approach has proved
its power in studying new important class of super--p--branes (such as
Dirichlet branes \cite{le} and the M-theory five--brane \cite{m5}) for which
supersymmetric equations of motion were obtained in the
geometrical approach \cite{hs1} earlier than complete supersymmetric actions
for them were constructed by standard methods \cite{c,m}. Thus, a problem
arises to relate the super--p--brane equations obtained from the action
with the field equations of the doubly supersymmetric geometrical
approach, and to convince oneself that they really describe one and the
same object. To accomplish this goal one should reformulate the action
principle for the super--p--branes such that it would yield the embedding
conditions of the geometrical approach in the most direct way. For
ordinary super--p--branes such an action has been proposed in \cite{bsv}.
The
construction is based on generalized action principle of the
group--manifold (or rheonomic) approach to superfield theories \cite{rheo}.
D. V. Volkov considered this approach as the most appropriate for
implementing geometry of the supersymmetric extended objects into the
description of their dynamics.

In this contribution we would like to review basic elements of the
generalized action construction and to show that it is also applicable to
the Dirichlet branes \cite{bst} and, at least partially, to the M--theory
five--brane (M--5--brane), thus allowing one to establish the relation
between the formulations of \cite{c,m} and \cite{hs1,hsw}.

On the way of reconstructing the super--p--brane actions we shall answer
another question connected with their $\k$--symmetry transformations,
namely, a puzzling fact that the $\k$--transformation of a ``kinetic" part
of the conventional super--p--brane actions is the integral of a
($p+1$)--form which compensates the $\k$--variations of a Wess--Zumino
term of the actions. This puzzle is resolved in a formulation where the
entire action of a super--p--brane is the integral of a differential
$(p+1)$--form in the worldvolume of the brane \cite{bpstv,bsv}.
To construct such an action one uses auxiliary harmonic
\cite{gikos} or twistor--like variables
which enable to get an irreducible realization of the
$\k$--transformations
(see \cite{stv,bzst,bzp} and references therein for superparticles,
superstrings and type $I$ super--p--branes). We shall also see that in
the case of the D--branes and the M--5--brane this version of the action
serves as a basis for the transition to a dual description of these
objects.

Consider the general structure of the action for a super--p--brane
propagating in a supergravity background of an appropriate space--time
dimension (which is specified by a brane scan \cite{scan}). We work with
actions of a Nambu--Goto (or Dirac--Born--Infeld) type that do not involve
auxiliary fields of intrinsic worldvolume geometry as in the Brink--Di
Vecchia--Howe--Tucker form \cite{bdht} of brane actions (see
\cite{c0,hull} for the BDHT approach to D--branes).

All known super--p--brane actions, except that of the M--5--brane which
contains a third term (see below), generically consist of two terms:
\begin{equation}\label{1}
~S=I_1~+~I_{WZ}=\int_{M_{p+1}}
d^{p+1}x e^{-{{p-3}\over
4}\phi}\sqrt{-\det{G_{mn}}}~+~\int_{M_{p+2}}W_{p+2}.
\end{equation}
The
symmetric part $g_{mn}$ of the matrix $G_{mn}\equiv g_{mn}+{\cal F}_{mn}$
in the first term of \p{1} describes a super--p--brane worldvolume metric
induced by embedding into a target superspace which is parametrized by
bosonic coordinates $X^{\underline m}(x)$ (${\underline m}=0,1,...,D-1$)
and fermionic coordinates $\Th^{\underline \mu}(x)$ $({\underline
\mu}=1,...,2^{[{D\over 2}]})$ collectively defined as $Z^{\underline
M}=(X^{\underline m},\Th^{\underline \mu})$. The worldvolume itself is
parametrized by small $x^m$ ($m=0,...,p$) with not underlined indices.
$\phi(Z)$ is a background dilaton field. Note that there is no such a
field in $D=11$ supergravity.

The antisymmetric part ${\cal F}_{mn}$ of $G_{mn}$, which is absent from
ordinary superbranes and nonzero for the D--branes and the M--5--brane,
contains the field strength of a gauge field propagating in the brane
worldvolume plus the worldvolume pullback of a
Grassmann--antisymmetric field of target--space supergravity.

In the case of the D--branes in $D=10$ the worldvolume
field is a vector field
$A_m(x)$ \cite{le,c}, the background field is a two--rank superfield
$B_{\underline{MN}}(X,\Th)$, and  ${\cal F}_{mn}$ has the form
\begin{equation}\label{Df}
{\cal F}^{(D)}_{mn}=e^{-{\phi(Z)\over 2}}(\partial_mA_n-\partial_nA_m
+\partial_mZ^{\underline N} \partial_nZ^{\underline
M}B_{\underline{MN}}).
\end{equation}

In the case of the M--5--brane the worldvolume gauge field is a
self--dual (or chiral) tensor field $A_{mn}(x)$,
and the background field is a three--rank
superfield $C_{\underline{LMN}}(X,\Th)$ of $D=11$ supergravity \cite{m5,m}.
The M--5--brane action also contains an auxiliary worldvolume scalar field
$a(x)$ \cite{pst} whose presence ensures manifest $d=6$ worldvolume
covariance of the model \cite{pst2,m}.
In this case the antisymmetric matrix takes
the form
\begin{equation}\label{Mf} {\cal
F}^{(M)}_{mn}={i\over{\sqrt{\partial_pa\partial^pa}}}
H^*_{mnl}\partial^la(x),
\qquad
H_{mnl}=6\partial_{[l}A_{mn]}+\partial_lZ^{\underline L}
\partial_mZ^{\underline N}\partial_nZ^{\underline M}C_{\underline{MNL}},
\end{equation}
where $*$ denotes Hodge operation, e.g.
$H^*_{mnl}={\sqrt{-g} \over 3!}
\varepsilon_{mnlpqr} H^{pqr}$.

\smallskip
The second term in \p{1} is a Wess--Zumino (WZ) term. Generically it is
more natural to define it as an integral of a closed
differential $(p+2)$--form over a $(p+2)$--dimensional manifold whose
boundary is the super--p--brane worldvolume. The structure of the WZ
term depends on the p--brane considered and (in general) includes
worldvolume pullbacks of antisymmetric gauge fields of
target--space supergravity and their duals (see \cite{c,m} for details).

The third term which one must add to the action \p{1} to describe the
M--5--brane dynamics is quadratic in $H_{mnl}$ \cite{m}:
\begin{equation}\label{i3}
I_3=\int d^6x{i\over{\sqrt{-\partial_pa\partial^pa}}}
{\cal F}^{(M)}_{mn}H^{mnl}\partial_la(x).
\end{equation}
In this case the action \p{1} plus \p{i3} is invariant under the local
symmetries
\cite{pst,pst2}
\begin{equation}\label{phi}
 \delta A_{mn}={{\varphi(x)}\over
{2(\partial a)^2}}(H_{mnp}\partial^pa-{\cal V}_{mn}),
\qquad
\delta a(x)={\varphi(x)}
\end{equation}
and
\begin{equation}\label{phi2}
\delta A_{mn}=\partial_{[m}a(x){\varphi_{n]}(x)},
\qquad
\delta a(x)=0
\end{equation}
where
$$
{\cal
V}^{mn}\equiv-2\sqrt{{{(\partial a)^2}\over{g}}}
{{\delta{\sqrt{-\det(G_{pq})}}}\over{\delta{\cal F}_{mn}}},
$$
and ${\varphi}$ and ${\varphi_m}$ are local gauge parameters.
The local symmetry \p{phi} allows one to gauge the field $a(x)$ away at the
expense of manifest Lorentz invariance of the M--5--brane action and the
local symmetry \p{phi2} is needed to ensure the self--duality condition
for $A_{mn}$.
These local symmetries are, in some sense, a bosonic analog of the fermionic
$\kappa$--symmetry (see below) whose gauge fixing also results in the loss
of Lorentz covariance.

The action \p{1} (plus \p{i3} in the case of the M--5--brane)
is invariant under the following $\kappa$--transformations of the
worldvolume fields
$$ i_{\kappa}E^{\underline\alpha} \equiv\delta_\kappa
Z^{\underline M}E_{\underline M}^{\underline\alpha}
=\kappa^{\underline\a}, \qquad
i_{\kappa}E^{\underline a}=0,
$$
\begin{equation}\label{k}
\delta_{\kappa}g_{mn}=-4iE_{\{m}\Gamma_{n\}}i_{\kappa}E,
\end{equation}
$$
\delta_{\kappa}{\cal F}^{(D)}=i_{\kappa}dB_{(2)},
\qquad
\delta_{\kappa}H=i_{\kappa}dC_{(3)}, \qquad \delta_\k a(x)=0,
$$
where
\begin{equation}\label{sv}
E^{\underline
A}=\left(dZ^{\underline M}E_{\underline M}^{\underline
a}(X,\Th),dZ^{\underline M}E_{\underline M}^{\underline\alpha}(X,\Th)\right)
\end{equation}
are target--space
supervielbeins pulled back into the worldvolume. They define
the induced metric
\begin{equation}\label{ig}
g_{mn}=\partial_mZ^{\underline M}
\partial_n Z^{\underline N}
E^{\underline{a}}_{\underline N} \eta_{\underline{a}\underline{b}}
E^{\underline{b}}_{\underline M},
\end{equation}
and $i_{\k}$ denotes the
contraction of the forms with the $\kappa$--variation of $Z^{\underline
M}$ as written above. The Grassmann parameter $\kappa^{\underline\a}(x)$
of the $\k$--transformations satisfies the condition
\begin{equation}\label{kg}
\kappa^{\underline\a}=\kappa^{\underline\beta}{\bar\Gamma}^
{~\underline\alpha}_{\underline\b},
\end{equation}
where ${\bar\Gamma}$ is a traceless matrix composed of the
worldvolume pullbacks of target--space Dirac matrices and the tensor
${\cal F}_{mn}$ such that $\bar\Gamma^2=1$.
The form of ${\bar\Gamma}$ is specific for a p--brane considered and
reflects the structure of the WZ term \cite{c,m}.

Eq. \p{kg} reads that not all
components (in fact only half) of $\k^{\underline\a}$ are independent,
which causes the (infinite) reducibility of the $\k$--transformations.
If one tries to get an irreducible set of $\k$--parameters in this
standard formulation, one should break manifest Lorentz invariance of the
models.
The geometrical approach considered below provides us with a
covariant way of describing independent $\k$--transformations.

For all super--p--branes the $\k$--variation \p{k} of the Wess--Zumino
term is (up to a total derivative) the integral of a $(p+1)$--form
\begin{equation}\label{wzv}
\delta_\k I_{WZ}=\int_{M_{(p+1)}}i_\k W_{(p+2)}.
\end{equation}
For the complete action to be $\k$--invariant the WZ variation must be
compensated by the variation of the NG or DBI-like term (and the term
\p{i3}). Thus, though these parts of the action are not the integrals of
differential forms, their $\k$--variations are. To explain this puzzling
fact it is natural to look for a formulation where the entire action is
the integral of a $(p+1)$--form.
When we deal with ordinary super--p--branes,
for which ${\cal F}_{mn}=0$, this can be easily done, since (apart from
the presence of the dilaton field) the NG term in \p{1} is the integral
volume of the world surface and can be written as the worldvolume
differential form integral
\begin{equation}\label{harm}
I_1=\int_{M_{(p+1)}}{1\over{(p+1)!}}E^{a_0}
\wedge E^{a_1} \wedge  ...  \wedge E^{a_p} \epsilon_{a_0 a_1... a_p},
\end{equation}
where $E^a=dx^mE_m^a(x)$ is a worldvolume vielbein form. Since we consider
induced geometry of the worldvolume, $E^a$ is
constructed as a linear combination of the target--space supervielbein
vector components \p{sv}
\begin{equation}\label{ind}
E^a=dx^m\partial_m Z^{\underline M}E_{\underline M}^{\underline b}
u_{\underline b}^{~a}(x).
\end{equation}
$u_{\underline b}^{~a}(x)$ are components (vector Lorentz harmonics
along the worldvolume) of an
$SO(1,D-1)$--valued matrix
\begin{equation}\label{u}
u_{\underline{b}}^{~\underline{\tilde a}}=(u_{\underline b}^{~a},
~u_{\underline b}^{~i})
\qquad a=0,...,p ~~i=p+1,...,D-1 ,
\end{equation}
\begin{equation}\label{ort}
u_{\underline a}^{~\underline{\tilde c}}
u_{\underline{\tilde c}}^{~\underline
b} = \delta^{~\underline b}_{\underline a}, \qquad
u_{\underline{\tilde a}}^{~\underline{c}}
u_{\underline{c}}^{~\underline{\tilde b}} =
\delta^{~\underline{\tilde b}}_{\underline{\tilde a}} =
diag(\d^b_a,\d^{ij}), \qquad \end{equation}
The orthogonality conditions
\p{ort} are invariant under the direct product of target--space local
Lorentz rotations $SO(1,D-1) \times SO(1,D-1)$ acting on $u$ from the left
and right, while the splitting \p{u} breaks one $SO(1,D-1)$ (tilded
indices) down to its $SO(1,p)\times SO(D-p-1)$ subgroup, which form a
natural gauge symmetry of the $p$--brane embedded into $D$--dimensional
space--time.

Surface theory tells us that \p{u} can always be chosen such that
on the world surface
\begin{equation}\label{ui}
E^i=dZ^{\underline M}E_{\underline M}^{\underline b}
u_{\underline b}^{~i}(x)\vert_{M_{p+1}}=0,
\end{equation}
i.e. orthogonal to the surface.

Dynamically one derives Eq. \p{ui} from the action \p{harm} by varying it
with respect to the auxiliary variables $u_{\underline b}^{~a}$
and taking into account the orthogonality condition \p{ort}.

In view of \p{ind}, \p{ui} and \p{ort} we see that the expression \p{ig}
for the induced metric reduces to $g_{mn}= E^{a}_mE_{an}$. Hence,
we can replace the determinant of $E^{a}_m$ written in \p{harm} with
$\sqrt{-\det{g_{mn}}}$ and return back to the NG form of the super-p-brane
action. This demonstrates the equivalence of the two formulations.

Note that only vector components $E^{\underline a}$ of
the target--space supervielbein \p{sv} enter the action \p{harm} through
Eqs. \p{ind}.  But in target superspace a supervielbein also has
components along spinor directions \p{sv} (i.e. $E^{\underline \a}$).
When
the supervielbein vector components undergo a local $SO(1,D-1)$
transformation with the matrix \p{u}, the supervielbein spinor components
are rotated by a corresponding matrix
$v_{\underline{\b}}^{~{\underline{\tilde\alpha}}}(x)$
of a spinor representation of the group $SO(1,D-1)$, the matrices
$u_{\underline b}^{~\underline{\tilde a}}$ and
$v_{\underline{\b}}^{~{\underline{\tilde \alpha}}}$ being related to
each other through the well--known formula
(see for instance \cite{bzst,bzp})
\begin{equation}\label{1b}
 u_{\underline{b}}^{~{\underline{\tilde a}}}
\Gamma^{\underline{b}}_{\underline{\alpha}\underline{\beta}} = \
v_{\underline{\a}}^{~\underline{{\tilde\d}}} ~
\Gamma^{\underline{\tilde a}}_{\underline{\tilde\d}\underline{\tilde\g}}
v^{~\underline{{\tilde\g}}}_{\underline{\b}}, \qquad
 u_{\underline{\tilde b}}^{~{\underline{a}}}
\Gamma^{\underline{\tilde
b}}_{\underline{\tilde\alpha}\underline{\tilde\beta}} =
u_{b}^{~{\underline{a}}}
\Gamma^{b}_{\underline{\tilde\alpha}\underline{\tilde\beta}} -
 u^{i{\underline{a}}}
\Gamma^{i}_{\underline{\tilde\alpha}\underline{\tilde\beta}} =
v_{\underline{\tilde\a}}^{~\underline{\d}} ~
\Gamma^{{a}}_{\underline{\d}\underline{\g}}
v^{~\underline{{\g}}}_{\underline{\tilde\b}}.
\end{equation}
The matrix $v_{\underline{\b}}^{~{\underline{\tilde\alpha}}}$ satisfies an
orthogonality condition analogous to \p{ort}. Thus, it is natural to
consider the spinor harmonic variables
$v_{\underline{\b}}^{~{\underline{\tilde\alpha}}}$ as independent and
$u_{\underline b}^{~\underline{\tilde a}}$ composed of the former.
The $SO(1,p)\times SO(D-p-1)$ invariant splitting of $v$
(analogous to \p{u}) is
\begin{equation}\label{v}
v_{\underline{\b}}^{~{\underline{\tilde\alpha}}}=
(v_{\underline{\b}}^{~\a q},
v_{\underline{\b}}^{~\dot{\a} \dot{q}}),
\end{equation}
where $\a , \dot{\a}$ are the indices of (the same or different)
 spinor representations of $SO(1,p)$
and $q, \dot{q}$ correspond to  representations of $SO(D-p-1)$.
The choice of
these representations depends on the dimension of the super--p--brane and
the target superspace considered and is such that
the dimension of the $SO(1,p)$ representations times the dimension of
the $SO(D-p-1)$ representations is equal to the spinor representation of
$SO(1,D-1)$.

To generalize Eq. \p{harm} to the case of the D-branes
one should take into account the presence of the antisymmetric
tensor ${\cal F}_{mn}$ in \p{1} as follows:
\begin{equation}\label{dir}
I_1 =
\int_{M_{(p+1)}}\Big({1\over{(p+1)!}}E^{a_0} \wedge E^{a_1}
\wedge  ...  \wedge E^{a_p}
\epsilon_{a_0 a_1... a_p} e^{-{p-3 \over 2} \phi} \sqrt{-det (\eta_{ab}
+
{\cal F}_{ab})}
\end{equation}
$$
+ Q_{p-1} \wedge [e^{-{1\over 2} \phi} (dA - B_{(2)})
-{1\over 2} E^b \wedge E^a {\cal F}_{ab}]\Big)~~ + ~~\int_{M_{p+2}}W_{p+2}
$$
where
we also included the WZ term, and the worldvolume scalar
${\cal F}_{ab}$ is an auxiliary antisymmetric tensor field with tangent
space (Lorentz group) indices and $Q_{p-1}$ is a Lagrange multiplier
differential form which
produces the algebraic equation
\begin{equation}\label{20}
{\cal F}^{(D)}_{2} \equiv 1/2 E^b \wedge E^a F_{ab} = e^{-{\phi \over
2}}(dA - B_{2}).
\end{equation}
Eq. \p{20} relates ${\cal F}_{ab}$ to the 2--form
${\cal F}^{(D)}_2$ of the original action \p{1}.

In the case of the M--5--brane its action \p{1} plus \p{i3}
is written as an integral of differential forms as follows
\begin{equation}\label{mf}
I_1+I_3=
\int_{M_{6}}\big({1\over{6!}}E^{a_0} \wedge E^{a_1}
\wedge  ...  \wedge E^{a_p}
\epsilon_{a_0 a_1... a_p} (\sqrt{-det (\eta_{ab}+{\cal F}_{ab})}-
{i\over{4\sqrt{-v_av^a}}}{\cal F}^{ab}H_{abc}v^c)
\end{equation}
$$
+ Q_{3} \wedge [dA_2 - C_{3}
-{1\over {3}} E^c\wedge E^b \wedge E^a H_{abc}]
$$
$$
+Q_5\wedge(da(x)-E^av_a\big)~~+~~\int_{M_{7}}W_{7}.
$$
In \p{mf}
${\cal F}^{ab}={i\over{\sqrt{-v_av^a}}}H^{*abc}v_c$, and
$H_{abc}(x)$ and $v_a(x)$ are auxiliary worldvolume scalar fields
which are expressed in terms of original fields $A_{mn}$ and $a(x)$
\p{Mf} upon solving the equations of motion for the Lagrange multiplier
forms $Q_{3}$ and $Q_{5}$.

The variation of the actions \p{dir} and \p{mf} with respect to the
auxiliary fields ${\cal F}_{ab}$, $H_{abc}(x)$ and $v_a(x)$ produces
algebraic expressions for the Lagrange multipliers $Q_{(n)}$, which
thus do not describe independent degrees of freedom of the models.
Note also that, at least for the Dirichlet branes with $p\le 4$, one can
invert the equations for $Q_{p-1}$ in terms of ${\cal F}_{ab}$, express
the latter in terms of the former and substitute them in the action.
This gives a dual worldvolume description of the D--branes
\cite{t,pst2,hull}.
Thus the actions \p{dir} and \p{mf} have the form which provides one with
a way to perform a dual transform of the superbrane models.

The worldvolume fields $Z^{\underline M}(x)$ and $A(x)$
(or ${\cal F}_{mn}$ and $H_{mnl}$) of the
super--p--branes are transformed under the $\k$--transformations as above
(see Eqs. \p{k}).

The $\k$--variation of the auxiliary fields and the
Lagrange multipliers can be easily obtained from their expressions
in terms of other fields whose $\k$--transformations are known.

To compute the $\k$--transformation of the actions \p{dir} and \p{mf}
we should also know the $\k$--variations of the Lorentz--harmonic fields
$u_{\underline b}^{~{\underline a}}$, which are genuine worldvolume
fields.  However these variations are multiplied by algebraic field
equations such as \p{ui} and \p{20} and, therefore, they can be
appropriately chosen to compensate possible terms proportional to the
algebraic equations that arise from the variation of other terms. It
means, in particular, that when computing $\delta_\k S$ we can freely use
these algebraic equations and, at the same time, drop the $\k$--variations
of these genuine worldvolume quantities if we are not interested in their
specific form.

Thus by construction the actions \p{dir} and \p{mf} are integrals of
$(p+1)$--forms ${\cal L}_{p+1}$ and one can show that their
$\k$--variation has the following general structure \cite{bst}:
\begin{equation}\label{25}
\delta S =
\int i_\k d {\cal L}_{p+1}=
\int i_\k\left(i{E}^{(-)}\gamma^{(p)}
{E}^{(-)}
- 2{E}^{(-)}
\gamma^{(p+1)}
\Delta\phi\right),
\end{equation}
where the second term is absent from the case of the M--5--brane,
\begin{equation}\label{31}
{E}^{(-){\underline\alpha}} = {1\over
2}{E}^{\underline\b}(1 - \bar\Gamma)^{~{\underline\alpha}}_{\underline\b},
\qquad \Delta_{{\underline{\a}}}\phi(X,\Th)=
{E}_{{\underline{\a}}}^{\underline M}\partial_{\underline M}\phi,
\end{equation}
$\gamma^{(p)}_{{\underline{\a}}{\underline{\b}}}$ and
$\gamma^{(p+1)}_{{\underline{\a}}{\underline{\b}}}$ are, respectively,
differential $p$--form and $p+1$--form constructed of
world\-volume--projected target--space gamma--matrices and the
tensor ${\cal F}_{mn}$ (see \cite{c,bst,62} for details).

The fact that $i_k d{\cal L} =0 $ and \p{25} is
$\kappa$--invariant follows from Eqs. \p{k} and \p{kg} which imply
$i_\k
{E}^{(-)\hat{\underline\alpha}} =0 = i_k E^{{\underline a}}$.

Note that, since the spinor parameter $\k$ corresponds to a
particular class of general variations of $\Th(x)$,
the knowledge of the $\k$--variation \p{25} of the
super--p--brane action enables one to directly get equations
of motion of $\Th(x)$ as differential form equations
\begin{equation}\label{6.9}
 i \gamma^{(p)}_{{\underline{\a}}{\underline{\b}}}
{E}^{(-){\underline{\b}}}
- ({1\over 2}(1-\bar{\Gamma})
\gamma^{(p+1)})_{{\underline{\a}} }
^{~{\underline{\b}} }~
\Delta_{{\underline{\b}}}\phi = 0.
\end{equation}

Let us now demonstrate how the presence of the Lorentz--harmonic fields
allows one to extract in a covariant way the independent parameters of the
$\k$--transformations (see \cite{bzst,bzp}
for ordinary super--p--branes)\footnote{An alternative
possibility of getting irreducible covariant $\kappa$--transformations
and their covariant gauge fixing has been discussed in \cite{k}.}.
For this we use the $SO(1,p)\times SO(D-p-1)$ decomposition of the spinor
harmonics \p{v}. To be concrete, consider the example of a Dirichlet
3--brane (p=3) in a background of type $IIB$ $D=10$ supergravity
\cite{bst}.  In this case the decomposition \p{v} of a $16\times 16$
matrix $v^{~\underline{{\tilde\alpha}}}_{\underline{\b}}$ takes the form
\begin{equation}\label{d3}
v^{~\underline{{\tilde\alpha}}}_{\underline{\b}} =
(v^{~\alpha }_{\underline{\alpha}q},
\bar{v}^{~\dot{\alpha}q}_{\underline{\alpha}} ),
 \end{equation}
where  $\a =1,2~,~ \dot{\a}=1,2$ are Weyl spinor indices of $SO(1,3)$,
$q=1,...,4$ are $SO(6)$ spinor indices and bar denotes complex
conjugation.

Using the exact form of the matrix $\bar\G$ and the
condition \p{kg} one can show \cite{bst} that the following 16 complex
conjugate components of the complex $\k$--parameter are independent:
\begin{equation}\label{kin}
\k^{\a}_q(x)=\k^{{\underline{\b}}}
v_{\underline{\b}q}
^{~\a}, \qquad
\bar{\k}^{\dot{\a}q}(x)= {\bar\k}^{{\underline{\b}}}
\bar v_{\underline{\b}}^{~{\dot\a} q}.
\end{equation}

By the use of independent parameters (such as \p{kin}) the
$\k$--transformations \p{k} can be rewritten in an irreducible form.
This realization of $\k$--symmetry is target--space covariant since
the parameters \p{kin} are target--space scalars and carry the indices
of the $SO(1,p)\times SO(D-p-1)$ group, the first factor of which is
identified with the Lorentz rotations in the tangent space of the
superbrane worldvolume and the second factor corresponds to an internal
local symmetry of the super--p--brane. Because of the fermionic nature
of these worldvolume $\k$--parameters it is tempting to treat them as the
parameters of $n$--extended local supersymmetry\footnote{The number $n$ of
worldvolume supersymmetries is such that $n\times
\dim{Spin(1,p)}=\dim{Spin(1,D-1)}$.} in the worldvolume of the
super--p--brane, and this was just a basic idea of \cite{stv}, which has
been fruitfully developed \cite{gs92}--\cite{berko} in the framework of
the doubly supersymmetric approach.

To make the local worldvolume supersymmetry manifest one should extend the
worldvolume to a world supersurface parametrized by $x^m$ and $n$
$SO(1,p)$--spinor variables $\eta^{\alpha q}$ all fields of the
super--p--brane models becoming worldvolume superfields.

Now, the differential form structure \p{dir} of super--p--brane actions
admits of an extension to worldvolume superspace by the use of generalized
action principles of the group--manifold (or rheonomic) approach
\cite{rheo} to supersymmetric field theories. This has been carried out
for the ordinary super--p--branes \cite{bsv} and the Dirichlet branes
\cite{bst}.  As to the M--5--brane, the presence of the term
$Q_5\wedge(da(x)-E^av_a\big)$ in \p{mf} causes problems to be solved yet
to lift the M--5--brane action to worldvolume superspace, since (without
some modification) this term would lead to rather strong (triviallizing)
restrictions on worldvolume supergeometry. Thus for the time being
further consideration is not applicable in full measure to the
M--5--brane action \p{mf}, though final superfield equations for the
superbranes which one gets as geometrical conditions of supersurface
embedding are valid for the M--5--brane as well. The relation of the
M--5--brane action \p{1} and \p{i3} \cite{m} and component field equations
of the M--5--brane obtained from the doubly supersymmetric geometrical
approach \cite{hs1,hsw} was established in \cite{62}.

The rheonomic approach exhibits in a vivid fashion geometrical properties
of supersymmetric theories, and when the construction of conventional
superfield actions for them fails the generalized action principle allows
one to get the superfield description of these models. As we shall sketch
below, in the case of super--p--branes the generalized action serves for
getting geometrical conditions of embedding world supersurfaces into
target superspaces, which completely determine the on--shell dynamics of
the superbranes \cite{bpstv,bsv,hs1,bst}.
However an open problem is how to extend the generalized action approach
to the quantum level.

Main points of this doubly supersymmetric construction are the following.

The generalized action for superbranes has the same form as \p{dir} but
with all fields and differential forms replaced with superforms in the
worldvolume superspace $\Sigma$=($x^m$, $\eta^{\alpha q}$).
The integral is
taken over an arbitrary $(p+1)$--dimensional bosonic surface ${\cal
M}_{p+1}=\left(x^m, \eta^{\alpha q}(x)\right)$ in the worldvolume superspace
$\Sigma$.
Thus, the surface ${\cal M}_{p+1}$ itself becomes a
dynamical variable, i.e. one should vary \p{dir} also with respect to
$\eta(x)$, however it turns out that this variation does not produce new
equations of motion in addition to the variation with respect to other
fields, the equations of motion of the latter having the same form as that
obtained from the component action we started with. But now
the fact that the surface ${\cal M}_{p+1}$ is arbitrary
and that the full set of such surfaces spans the whole worldvolume
superspace makes it possible to consider these
equations of motion as equations for the superforms and
superfields defined in the whole worldvolume superspace $\Sigma$.
The basic superfield equations thus obtained are Eqs. \p{ui} and \p{6.9}
(note that now the external differential also includes the
$\eta$--derivative).
 Eqs. \p{ui} and \p{6.9} tell us
that induced worldvolume supervielbeins $\left(e^a(x,\eta), e^{\a q}(x,
\eta)\right)$ can be chosen as a linear combination of
$E^{\underline b}u_{\underline b}^{~a}\equiv E^a$ and
$E^{\underline \b}v_{\underline \b}^{~\a q}\equiv E^{\a q}$
\cite{bpstv,hs1,bst}:
\begin{equation}\label{6.12}
e^a  = E^b(m^{-1})_b^{~ a}  \qquad \Rightarrow
\qquad E^{~a}_{{\a}q} = 0,
\end{equation}
\begin{equation}\label{6.20'}
e^{\a q} =E^{\a q} + E^a \chi_{a}^{~\a q}(x,\eta),
\end{equation}
as well as that
\begin{equation}
\label{fermi}
E^{(-)\underline{\a}}_{\a q} \equiv
\Big(E_{\a q}{1 \over 2} (1-\bar{\G})\Big)^{\underline{\a}}  = 0 ,
\end{equation}
The choice of
the matrix $m_b^{~a}(x,\eta)$ is a matter of convenience and can be used
to get the main spinor--spinor component of the worldvolume torsion
constraints in the standard form $T^a_{\a q, \b r}=i\d_{qr}\g^a_{\a\b}$.
In this case $m_b^{~a}$ is constructed out of worldvolume gauge fields
\cite{hs1,bst,62}.

Eq. \p{6.12} together with eq. \p{ui} implies the basic
geometrodynamical condition
\begin{equation}\label{6.21} E_{\a q}^{ \underline a} = 0
\end{equation}
which in the doubly supersymmetric approach to super--p--branes determines
the embedding of the worldvolume superspace into the target superspace.
In many interesting cases such as $D=10$ type II superstrings
\cite{gs2,bpstv}
and D--branes \cite{hs1,bst}, and $D=11$ branes \cite{bpstv,hs1}
the integrability conditions for \p{ui}, \p{6.12} and \p{6.20'}
reproduce all the equations of motion of these extended objects and also
lead to torsion constraints on worldvolume supergravity \cite{bsv,bst}.

Note that for the D--branes and the M--5--brane the embedding conditions
analogous to \p{ui}, \p{6.12}, \p{6.20'} and \p{fermi}
were initially not
derived from an action, which was not known at that time, but
postulated \cite{hs1} on the base of the previous knowledge of analogous
conditions for ordinary super--p--branes \cite{bpstv,bsv}.

To conclude, we have demonstrated how the super--p--brane
action can be reconstructed as the worldvolume integral of a
differential $(p+1)$--form. The use of the Lorentz--harmonic variables
in this formulation makes the $\k$--symmetry transformations
to be performed with an
irreducible set of fermionic parameters being worldvolume spinors.
This indicates that the $\k$--symmetry originates from extended local
supersymmetry in the worldvolume. We have shown how this worldvolume
supersymmetry becomes manifest in a worldvolume superfield
generalization of the super--p--brane action. The superfield equations
derived from the latter are the geometrical conditions of embedding
worldvolume supersurfaces swept by the superbranes in target superspaces.
Thus, the approach reviewed in this article serves as a bridge between
different formulations developed for describing superbrane dynamics.

{\bf Acknowledgements.} We would like to thank our collaborators Kurt
Lechner and Alexei Nurmagambetov with whom we obtained many results
reported herein. Work of P.P. and M.T. was supported by the European
Commission TMR programme ERBFMRX--CT96--0045 to which  P.P. and M.T.
are associated.
I.B. thanks Prof. M. Virasoro for hospitality at the ICTP.
I.B., and D.S. acknowledge partial support from grants of
the Ministry of Science and Technology of Ukraine and the INTAS Grants N
93--127--ext, N 93--493--ext and N 93--0633--ext.

\end{document}